\documentclass[aps,prb,preprint,superscriptaddress]{revtex4-1}

\usepackage{graphicx}
\usepackage{dcolumn}
\newcommand{\didv}{\ensuremath{\mathrm{d}I/\mathrm{d}V}}
\begin{document}

\title{On local sensing of spin Hall effect in tungsten films by using STM based measurements}

\author{Ting Xie}
	\email[]{tingxie@terpmail.umd.edu}
	\affiliation{Department of Electrical and Computer Engineering, University of Maryland, College Park, Maryland 20742, USA}
\author{Michael Dreyer}
	\affiliation{Department of Physics, University of Maryland, College Park, Maryland 20742, USA}
\author{David Bowen}
\author{Dan Hinkel}
\author{R. E. Butera}
\author{Charles Krafft}
	\affiliation{Laboratory for Physical Sciences, College Park, Maryland, 20740, USA}
\author{Isaak Mayergoyz}
 \affiliation{Department of Electrical and Computer Engineering, University of Maryland, College Park, Maryland 20742, USA}

\date{\today}

\begin{abstract}
The spin Hall effect in tungsten films has been experimentally studied by using STM-based measurements. These measurements have been performed by using tungsten and iron coated tungsten tips. In the case of tungsten tips, it has been observed that the current flow through the tungsten film results in an appreciable asymmetry in the tunneling current with respect to the change in the polarity of the tunneling voltage. It is reasoned that the cause of this asymmetry is the accumulation of spin polarized electrons on the tungsten film surface due to the spin Hall effect. This asymmetry is not affected by the change of the direction of the bias current through the film. However, in the case of iron coated tungsten tips, it has been observed that a change in the direction of the bias current does lead to an additional asymmetry in the tunneling current. It is thus experimentally demonstrated that this asymmetry is caused by the SHE and spin-dependent density of states of iron-coated tips. 
\end{abstract}


\maketitle

\section{Introduction}
The spin Hall effect (SHE) has recently attracted much attention due to its general theoretical interest as well as its potential technological applications in the field of spintronics \cite{Sinova}. The manifestation of the SHE is the accumulation of spin-polarized electrons at the boundaries of current-carrying samples \cite{Dyakonov, Hirsch}. The strength of the SHE is characterized by the spin Hall angle (SHA), which has been reported to be large in metals such as platinum, tantalum and tungsten \cite{Hoffmann}. The surface accumulation of spin-polarized electrons may cause (through diffusion) the injection of spin current without the injection of charge flow into adjacent metallic layers. This promises the use of the SHE for the development of novel spintronic devices. In the literature, experimental studies of the SHE in semiconductors and metals have been performed by using optical measurements \cite{Kato, Erve} as well as spin-torque phenomena \cite{Liu, Pai, Parkin}. In this paper, we demonstrate for the first time the sensing of spin dependent electron tunneling arising from the SHE in tungsten films by using scanning tunneling microscopy (STM) based measurements. The local sensitivity of the nanometer-scale STM tunneling junction and the ability to detect the accumulation of spin polarized electrons on the surface offers the possibility of identifying spatial variations in the strength of the SHE in current-carrying samples at the nanoscale. This may lead to a better understanding of the SHE. 

We performed the STM based measurements of the SHE with conventional tungsten tips as well as iron-coated tungsten tips. In the case of tungsten tips, we observed an asymmetry in the tunneling currents for opposite polarity, but identical magnitude, of the tunneling voltage in the presence of a bias current flow through tungsten films. By precluding other possible causes of the asymmetry such as thermal expansion and thermionic emission, it is argued that this asymmetry in the tunneling currents is caused by the asymmetry in the tunneling process. The latter is due to the presence of spin-polarized electrons at the tungsten film surface induced by the SHE. The accumulation of spin-polarized electrons produced by bias currents is further demonstrated by using an iron-coated tungsten tip. In this case, we have observed an additional tunneling current asymmetry caused by a change in the direction of the bias current flow through tungsten films. This asymmetry reflects the change of the orientation of the electron spin polarization at the tungsten surface as well as the spin dependent density of state of iron-coated tips. Thus, the obtained results clearly demonstrate the presence of bias-current-induced spin-polarized electrons at the tungsten film surface, which is the unique manifestation of the SHE.

\section{Experimental Details}
\subsection{Tungsten film preparation and characterization}
\begin{figure}
\includegraphics[width=1\columnwidth]{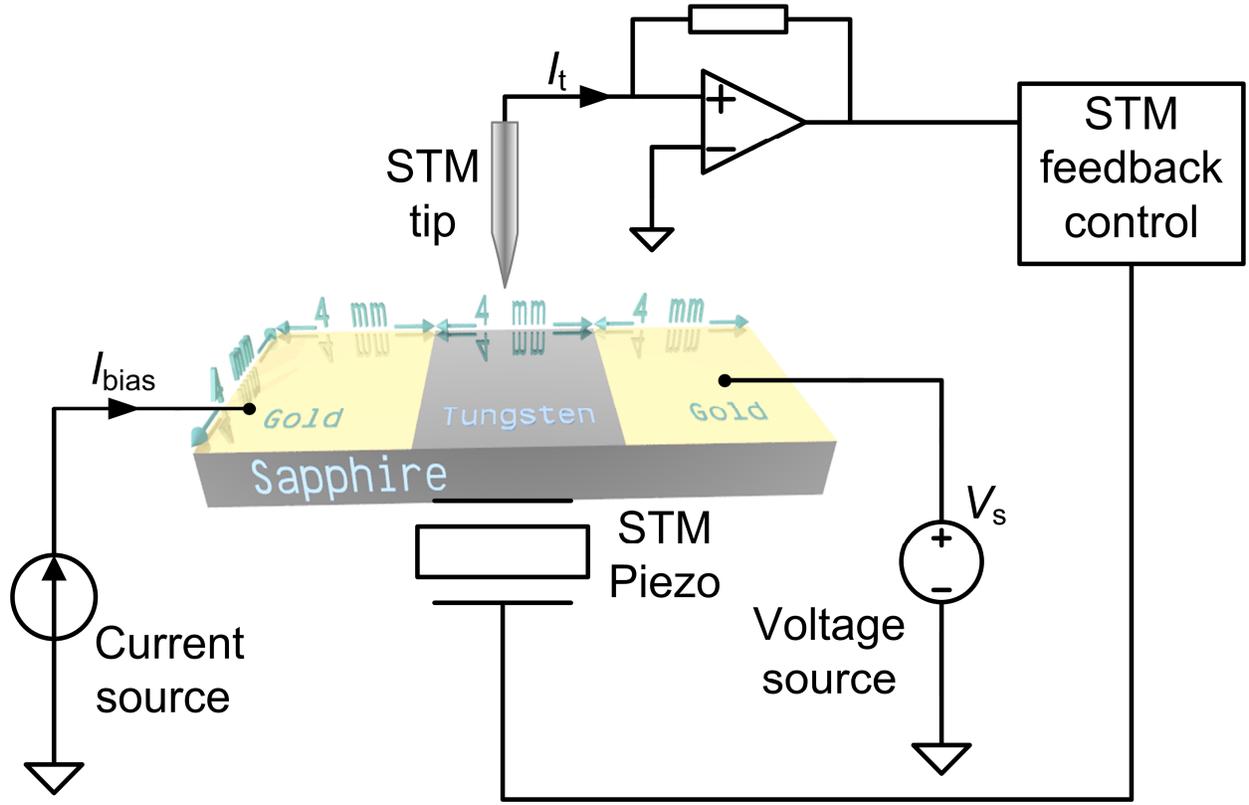}%
\caption{\label{fig1} Schematic of fabricated tungsten sample and the electrical connections of the experimental setup.}
\end{figure}

We used a two-chamber Omicron ultra-high vacuum (UHV) system to fabricate and analyze tungsten films \textit{in-situ}. The tungsten films were deposited on 12 mm $\times$ 4 mm $\times$  0.4 mm sapphire substrates by using a DC magnetron sputtering technique at a base pressure below $1 \times 10^{-9}$ Torr. Gold pads (4 mm $\times$ 4 mm $\times$ 1 $\mu$m) were pre-patterned through a shadow mask onto each end of the substrates to ensure good electrical contact before the tungsten films were deposited. During the tungsten deposition process, the argon sputtering pressure and the sputtering power were maintained at 3.0 mTorr and 6 W, respectively. These parameters were chosen in accordance with prior reports \cite{Pai, Xiao} to achieve the $\beta$-phase of tungsten films which exhibits a pronounced SHE. The length (i.e. distance between the gold pads) and the width of the tungsten films were ~4 mm each (see Fig.~\ref{fig1}). We studied the SHE in four tungsten film samples (see Table~\ref{table1}). The thickness of each tungsten film was controlled by the deposition time at a pre-calibrated deposition rate of 0.95 nm/min. The tungsten films were then transferred to the STM chamber for topographic characterization and for the subsequent STM study of the SHE. STM images of the fabricated tungsten films can be found in the Supplemental Material Fig. S1 \cite{Supplemental}. The film resistivities were measured to be $250 \pm 30$ $\mu\Omega\cdot$cm. These resistivities agree well with those reported for β-phase tungsten films in the literature \cite{Pai, Xiao}. We used electrochemically-etched tungsten tips which were e-beam heated under UHV conditions to remove the native oxides before loading them into the STM stage. Iron-coated tungsten tips (with iron layer thickness of $\sim18$ nm) were produced using an e-beam deposition technique in UHV after the oxide removal. During all tunneling measurements the system pressure was kept below $1 \times 10^{-10}$ Torr. 
\begin{table}[b]
\caption{\label{table1}%
The thickness of fabricated tungsten films.
}
\begin{ruledtabular}
\begin{tabular}{ccccc}
\textrm{Sample}& \textrm{F1}& \textrm{F2}& \textrm{F3}& \textrm{F4}\\
\colrule
\textrm{Thickness (nm)}& 5& 5& 7& 40\\
\end{tabular}
\end{ruledtabular}
\end{table}

\subsection{Experimental setup for local sensing of the SHE}
To perform the local study of the SHE, we modified the Omicron STM system in order to maintain a constant tunneling voltage in the presence of a bias current flow through the tungsten film as shown in Fig.~\ref{fig1}. The tungsten film sample is connected to a current source and a voltage source at its left and right electrical contacts, respectively. The controllable current source provides the desired bias current through the sample while the voltage source is used to apply a desired tunneling voltage $V_{g}$ between the STM tip and the sample. The bias current $I_{bias}$ creates a voltage drop along the resistive tungsten film which affects the tunneling voltage. To resolve this issue, we developed a potentiometry technique to properly compensate the bias-current-induced voltage $V(x, y)$ between the tunneling location and the right terminal of the sample \cite{Xie}. This technique can be briefly described as follows. The bias-current-induced surface potential $V(x, y)$ is determined by applying Kirchhoff’s voltage law (KVL) to the loop formed by the ground, voltage source, part of the conducting sample, tunneling gap, tip, pre-amplifier and ground (this loop is indicated by the blue dashed line in the Supplementary Material Fig. S2a \cite{Supplemental}). Since the STM tip is virtually grounded by the pre-amplifier, the KVL equation for the above loop can be written as follows
	\begin{equation}
	\label{eq1}
	V_{s} +V(x, y) = V_{g} + I_{t}{\cdot}R_{tip},
	\end{equation}
where $V_{s}$ is the source voltage, $V_{g}$ is the tunneling voltage, $I_{t}$ is the tunneling current, while $R_{tip}$ is the STM tip resistance. It is apparent that zero tunneling current corresponds to zero tunneling voltage. Therefore, the right-hand side of Eq.~\ref{eq1} is equal to zero when $I_{t} = 0$. In this case, the surface potential $V(x, y) = -V_{s}(0)$, where $V_{s}(0)$ is the voltage value at which $I_{t} = 0$. Thus, this value can be found by locating the intersection of the $I_{t}-V_{s}$ curve with the $I_{t} = 0$ line. Typical $I_{t}-V_{s}$ curves and the corresponding values of $V(x, y)$ obtained at $I_{bias} = 0$ mA, 0.2 mA and -0.1 mA are shown in Supplemental Material Fig. S2b. In the STM study of the SHE, it is very important to maintain a desired tunneling voltage between the tip and the conducting film in the presence of current flow through the film. This can be achieved by neglecting the very small term $I_{t}{\cdot}R_{tip}$ (~nV) in Eq.~\ref{eq1} and by setting the source voltage $V_{s} = V_{g}-V(x, y)$. By using this technique, the tunneling voltage can be directly controlled by the voltage source for any value of the bias current through the film. Namely, one can apply a desired voltage across the tunneling gap at any position of the tip with respect to the film and for any value and direction of the bias current. 

\section{Results}
\begin{figure}
\raisebox{0.12\columnwidth}{\includegraphics[width=0.49\columnwidth]{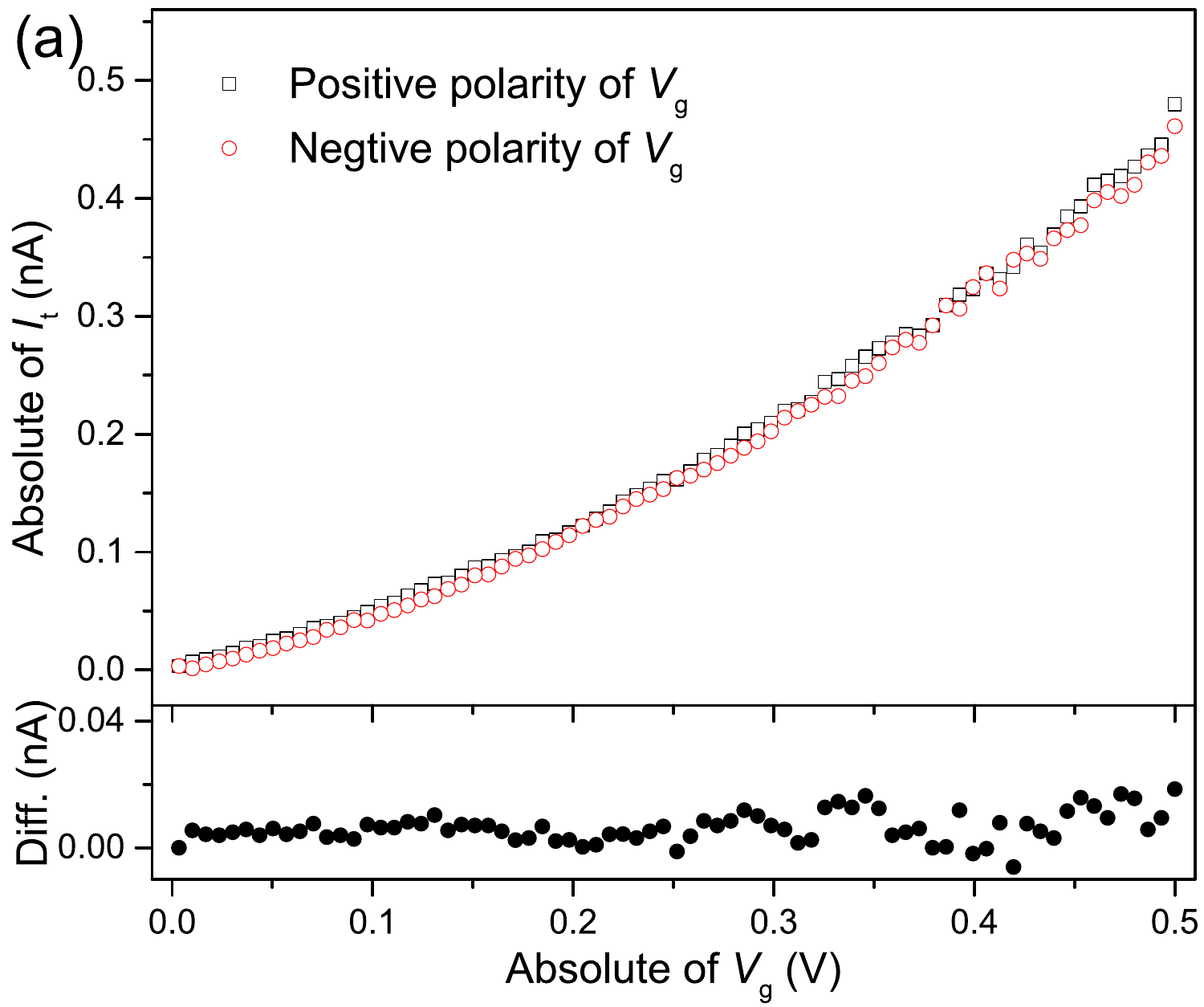}}\hfill
\includegraphics[width=0.49\columnwidth]{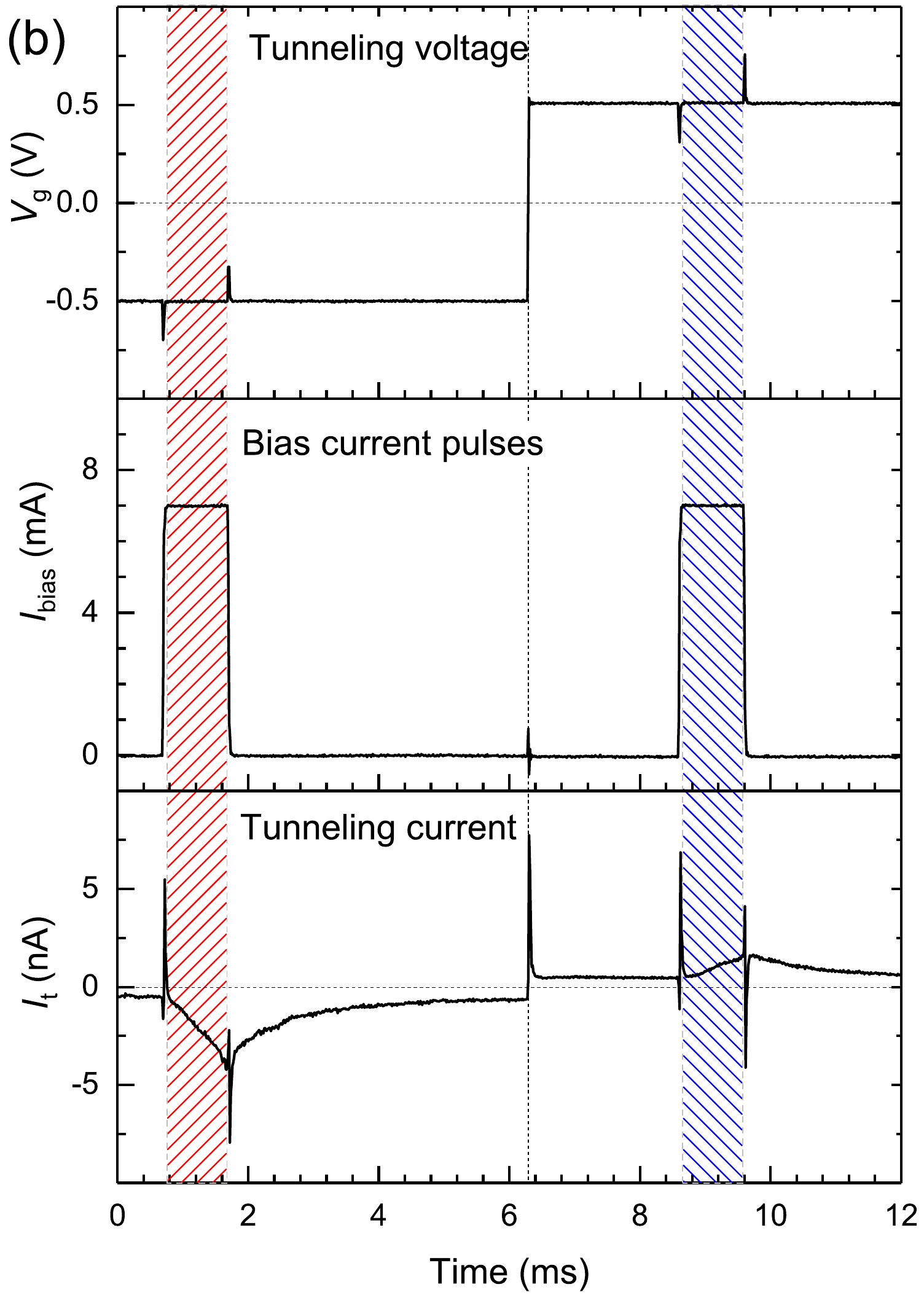}%
\caption{\label{fig2} (a) Magnitude of tunneling current vs. magnitude of tunneling voltage with no bias current applied. The lower graph shows the difference in the magnitude of tunneling currents measured at opposite polarities of the tunneling voltage. (b) Time variations of tunneling voltage, bias current and tunneling current in STM based SHE measurements with the STM feedback control turned off. It is apparent that there is an asymmetry in the tunneling current during the bias current pulses with respect to the change in polarity of the tunneling voltage (see red and blue shaded areas).}
\end{figure}

Figure~\ref{fig2}a presents a typical measurement of the tunneling current as a function of the tunneling voltage between film F1 and a tungsten tip in the absence of a bias current flow through the film. During the measurement, the feedback control of the STM was kept off. The red and black symbols represent the magnitude of the tunneling current measured at positive and negative polarities of the tunneling voltage, respectively, and the lower graph reveals their difference. It can be concluded from this figure that the magnitude of the tunneling current is nearly (even) symmetric with respect to a change in the polarity of the tunneling voltage, which is to be expected for a tungsten-tungsten tunneling junction. Then an extensive study of tunneling currents in the presence of bias currents through film F1 at different polarities of tunneling voltages were conducted while still using a tungsten tip. The bias-current-induced voltage drop in the tungsten film was compensated by using the above described potentiometry method to maintain a desired voltage across the tunneling gap while changing the bias current through the film. Figure~\ref{fig2}b shows time evolutions of the tunneling voltage, the bias current, and the tunneling current obtained from one SHE measurement on sample F1. In the SHE experiment, in order to establish a desired tunneling gap, a measurement starts by stabilizing the tip at a given tunneling current and tunneling voltage polarity. Then, the time evolution of the tunneling current is measured while applying a current pulse through the film. Here, bias current pulses of $\sim1$ ms in duration were used to minimize Joule heating effects which are discussed below. The choice of this pulse duration was due to bandwidth limitations of the STM electronics. After some time interval, the tunneling voltage polarity is changed and the time evolution of the tunneling current is measured again with an identical current pulse through the film. During the entire measurement, the feedback of the STM piezo was turned off. In the measurement demonstrated in Fig.~\ref{fig2}b, the tunneling voltage was switched between -0.5 V and 0.5 V and the amplitude of the bias current pulses were kept at 7 mA. The voltage polarity and the corresponding electron tunneling direction are indicated in the inset of Fig.~\ref{fig2}b.

It is evident from Fig.~\ref{fig2}b that the measured tunneling current exhibits spikes at the edges of the bias current pulses. These spikes can be attributed to displacement currents caused by fast transient variation in the tunneling voltage across the capacitive tip-sample junction at pulse edges. The widths and amplitudes of the spikes can be controlled by the ramping rate of the current pulses (see Supplemental Material Fig. S3 \cite{Supplemental}). During the bias current pulse, two effects were observed: a monotonic increase of the tunneling current in time, and an appreciable asymmetry in the tunneling current with respect to the change in tunneling voltage polarity. The temporal increase in the tunneling current can be attributed to a reduction of the tip-sample distance caused by thermal expansion of the sample during the bias current pulse. In fact, this reduction was confirmed by an atomic force microscopy study performed on the tungsten film under the same conditions (see Supplemental Material Fig. S4 for more explanation \cite{Supplemental}). The tunneling current asymmetry is remarkable because it contrasts with the practically symmetric nature of the tunneling current in the absence of the bias current flow through the film (see Fig.~\ref{fig2}a).

\begin{figure}
\includegraphics[width=0.49\columnwidth]{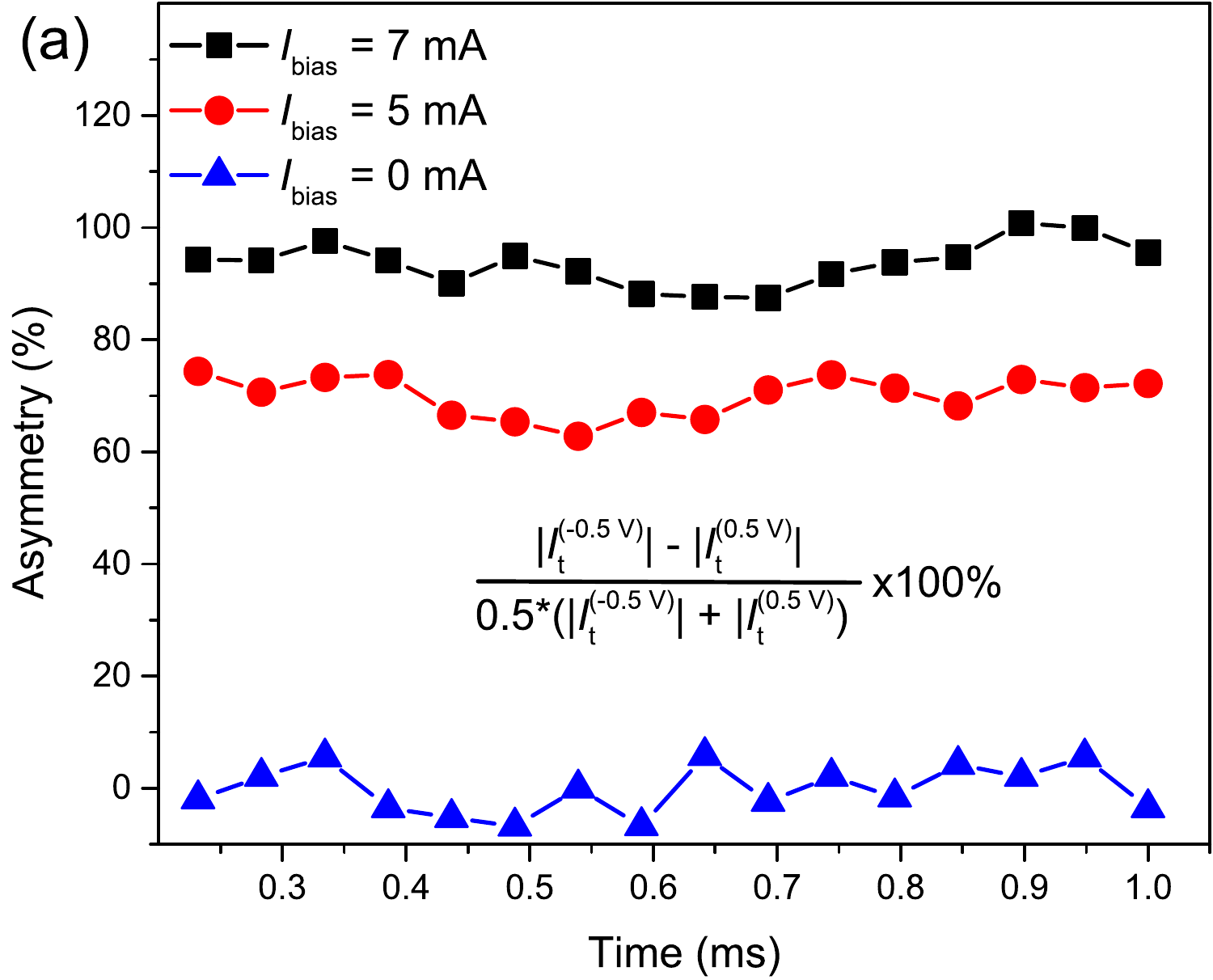}\hfill%
\includegraphics[width=0.49\columnwidth]{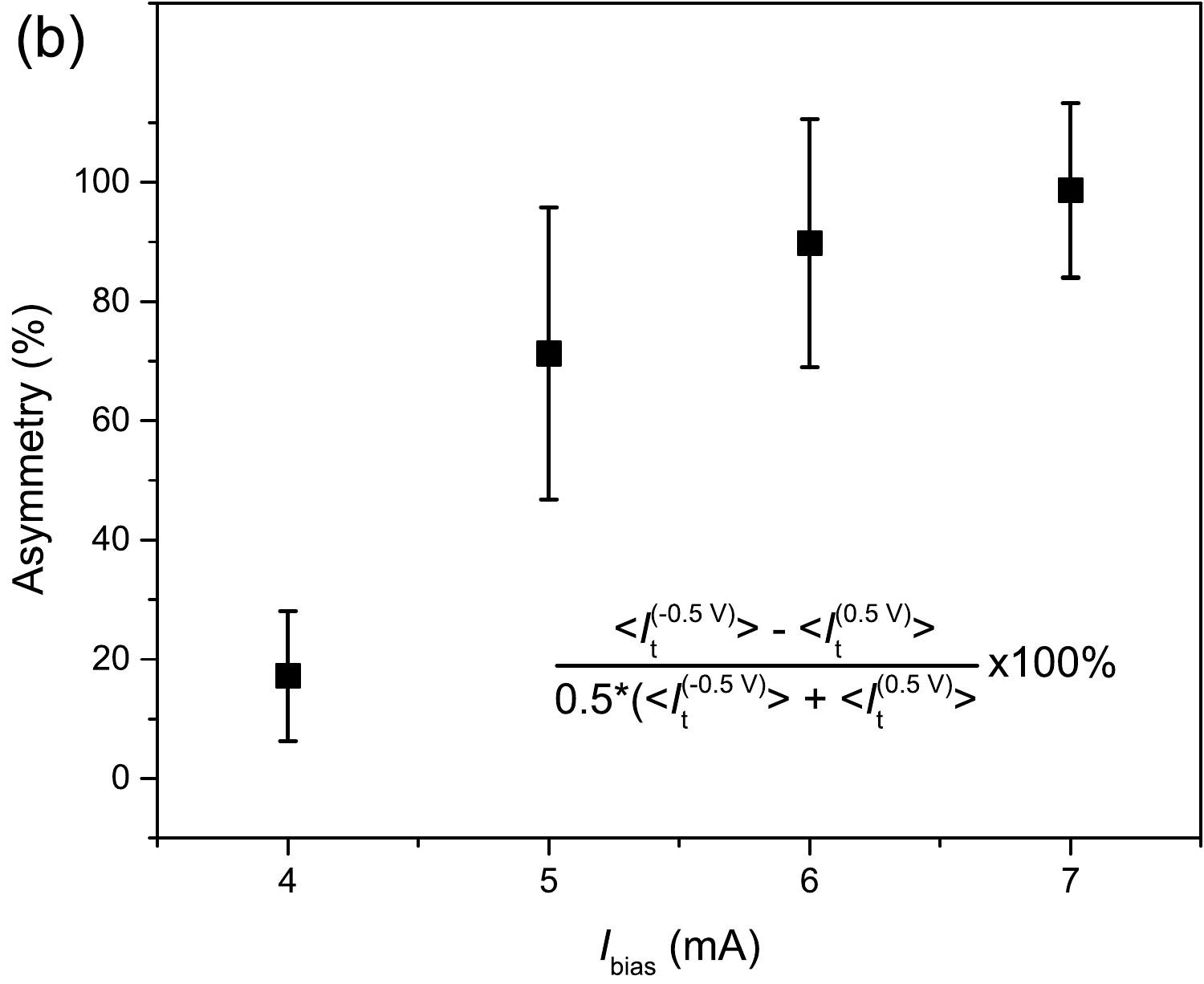}\vspace{8pt}
\includegraphics[width=0.49\columnwidth]{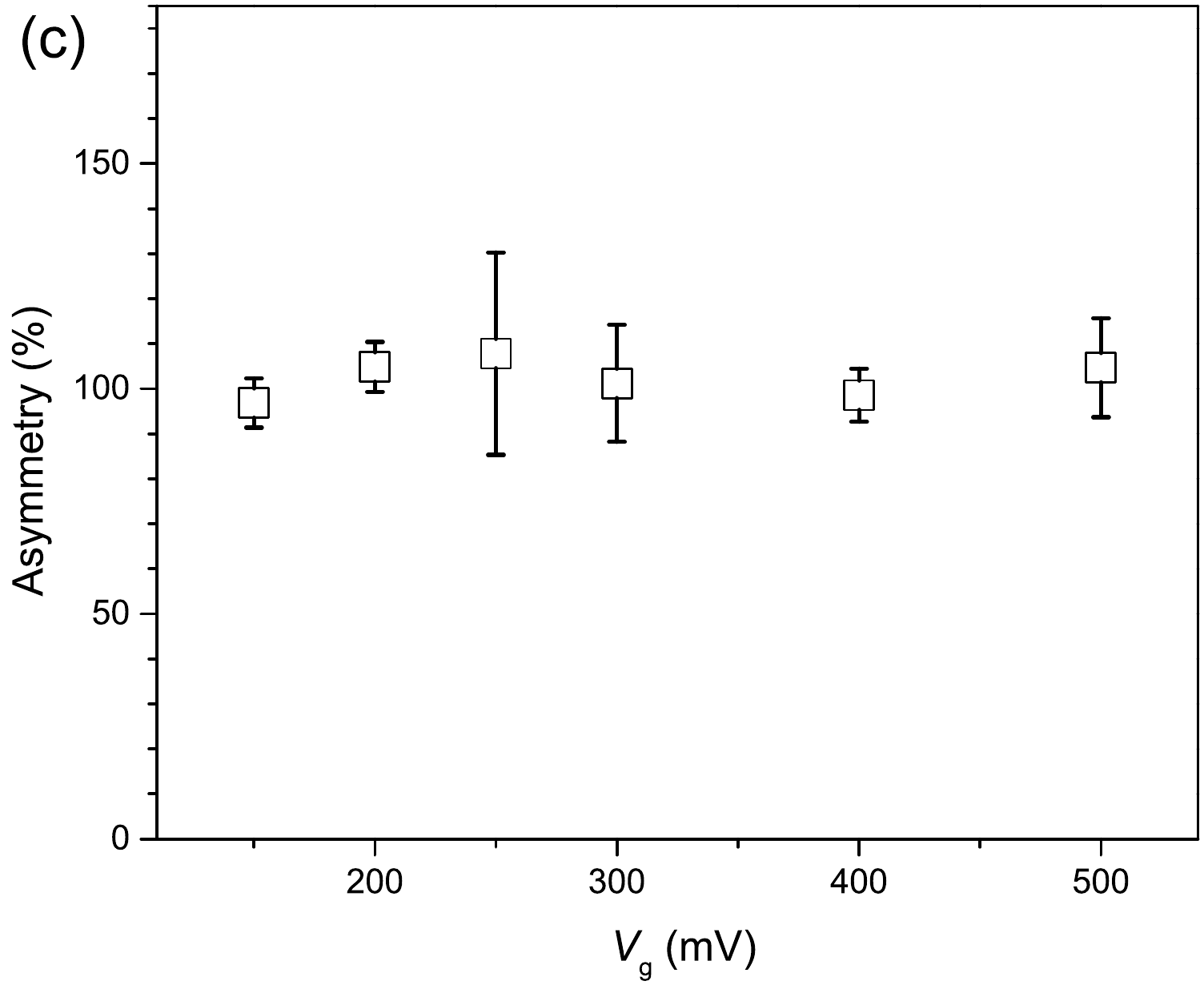}%
\caption{\label{fig3} (a) Asymmetry of tunneling currents measured at 0.5 V and -0.5 V tunneling voltages in the presence of bias current pulses with respect to time. The asymmetry is calculated by the equation shown in the figure, where $|I_{t}^{(-0.5 V)}|$ and $|I_{t}^{(0.5 V)}|$ are the magnitudes of tunneling currents measured at tunneling voltages of -0.5 V and 0.5 V, respectively. (b) and (c) Measured tunneling current asymmetry (time-averaged over the bias current pulse period) as a function of the magnitude of bias current and tunneling voltage, respectively. Error bars represent deviations observed for multiple measurements under the same conditions.}
\end{figure}

To more clearly illustrate the asymmetry due to the tunneling voltage polarity change, the normalized difference between the tunneling currents at corresponding instances of time with respect to rising edges of the bias current pulses (7 mA, 5 mA, and 0 mA) are plotted in Fig.~\ref{fig3}a. The normalization was performed by dividing the differences in the tunneling currents by their mean values. As a result of this normalization, the effects of the temporal increases in the tunnel current due to the tip-sample distance changes are eliminated. Remarkably, the normalized asymmetry values remain practically constant throughout the duration of the bias current pulse. In the absence of a bias current (see blue triangles in Fig.~\ref{fig3}a), the normalized asymmetry was near zero, which agrees with Fig.~\ref{fig2}a. In contrast, in the case of 5 mA and 7 mA bias current pulses, substantial asymmetries in the tunneling currents for opposite polarities of the tunneling voltage were observed (see red circles and black squares in Fig.~\ref{fig3}a). Multiple measurements at various values of bias current for film F1 were performed to reveal the dependence of the tunneling current asymmetry on the bias current amplitude. As shown in Fig.~\ref{fig3}b, the asymmetry increases with the bias current from 4 mA to 7 mA.

Several possible sources of the tunneling current asymmetry were considered, including thermal expansion of the sample, thermionic emission from the film, as well as the accumulation of spin-polarized electrons at the film surface due to the SHE. We determined that thermal expansion cannot result in the observed tunneling current asymmetry because the current pulses are identical for the two tunneling voltage polarities, therefore the thermal effect would be identical. Additionally, the asymmetry remained when the sequence of the tunneling voltage polarity was changed (see Supplemental Material Fig. S5a \cite{Supplemental}). This implies that any possible asymmetry due to a change in the tunneling gap resulting from a residual thermal expansion is precluded. As far as the thermionic emission is concerned, calculations show that the emission current from the tungsten film is negligible ($~10^{-17}$ A/$\mathrm{cm^2}$).\cite{Murphy} By precluding other possibilities, it can be then reasoned that the measured tunneling current asymmetry is caused by the accumulation of spin-polarized electrons (due to the SHE). The latter alters the electron tunneling process with respect to the polarity change of the tunneling voltage. Indeed, there is asymmetry in the tunneling process. Tunneling from the tungsten tip to the tungsten film involves non-spin polarized electrons, while changing the tunneling voltage polarity, tunneling of spin-polarized electrons from the tungsten film to the tungsten tip is realized. It is worthwhile to note that spin-dependent tunneling through potential barriers in semiconductors were theoretically studied in the literature \cite{Perel2003,Perel2004,Wang}, where it was demonstrated that the spin polarization of electrons may appreciably affect the tunneling process through the barrier due to the spin-orbit coupling (SOC). 

The SHE as a cause of the asymmetry is consistent with its increase with the increase in the bias current as shown in Fig.~\ref{fig3}b. This is because the accumulation of spin-polarized electrons is proportional to the total number of injected electrons \cite{Hoffmann}. It was also observed that the asymmetry may change its magnitude at different locations (see Supplemental Material Fig. S5b \cite{Supplemental}). Rarely, at some locations, the asymmetry was reduced to zero and even changed its sign. However, at these locations no relationship between the negative asymmetry and the magnitude of the bias current was observed. The variations in the asymmetry may be attributed to highly non-uniform (on the nanoscale) distribution of the bias current density due to the nanoscale granular surface structure of fabricated tungsten films. We also studied the tunneling current asymmetry as a function of the tunneling voltage magnitude. The results, as presented in Fig.~\ref{fig3}c, reveal a nearly constant asymmetry for a tunneling voltage variation between 150 mV and 500 mV. 

\begin{figure}
\includegraphics[width=0.7\columnwidth]{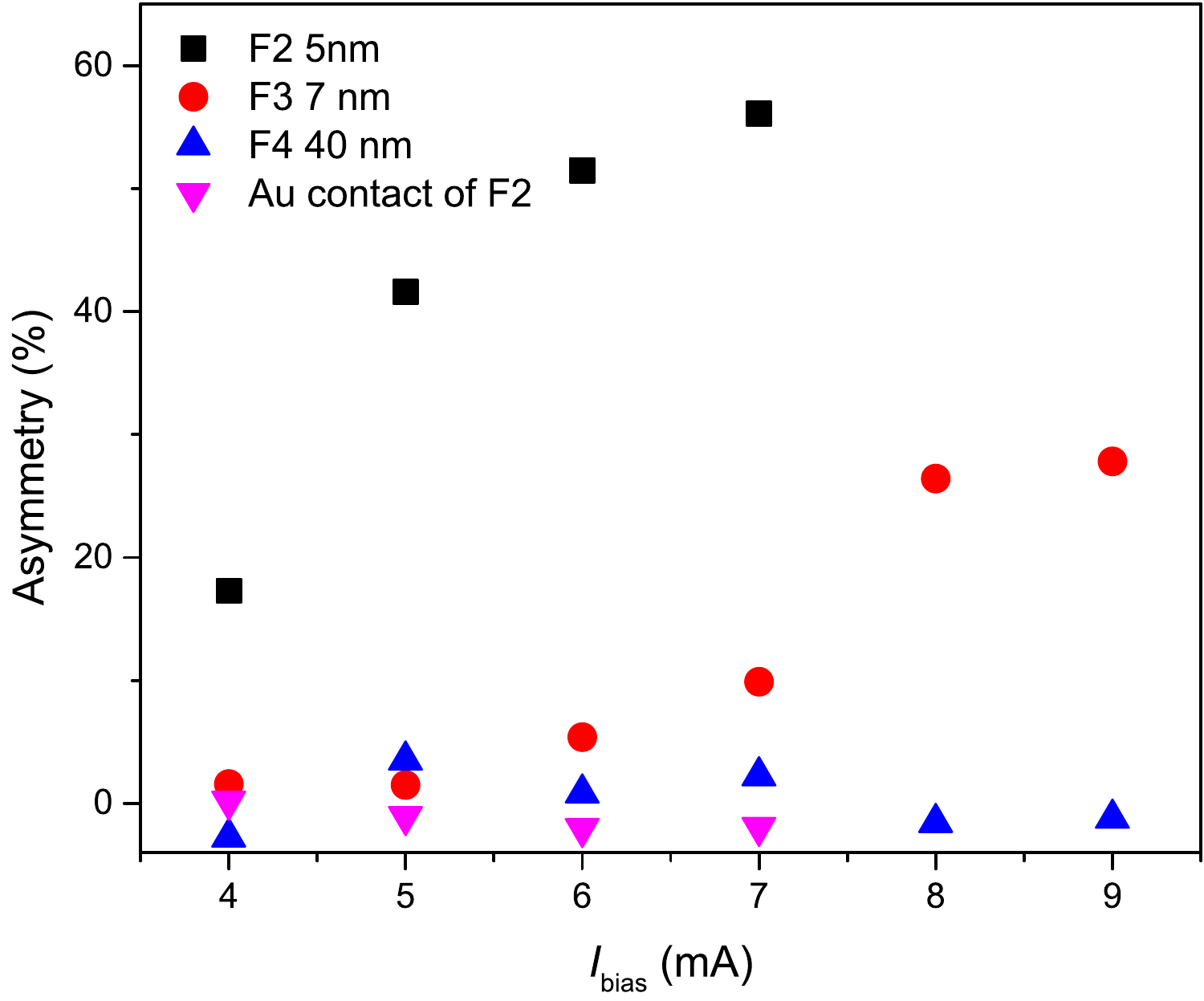}%
\caption{\label{fig4} Tunneling current asymmetry as a function of the bias current for tungsten films F2, F3, and F4, respectively, as well as the asymmetry measured at the gold contact of F2.}
\end{figure}

For comparison purposes, the asymmetry measurements were repeated on three other tungsten film samples, as well as on the gold contacts of sample F2. The results of these measurements are presented in Fig.~\ref{fig4}. It is clear from this figure that for the same bias currents the tunneling current asymmetry decreases as the film thickness increases. This can be explained by the reduction of the applied electric field in thicker tungsten films, which results in the reduction of the SOC \cite{Dresselhaus} in films and consequently the SOC induced SHE \cite{Sinova2015}. A similar argument is applicable to the gold contact area which has a much higher conductivity and thickness than the tungsten films and thus exhibits a significantly reduced electric field. The measurement on the gold contact of the sample F2 works as a good control experiment, and the disappearance of the asymmetry in the gold area precludes any possible experimental flaws as causes of the measured asymmetry at the tungsten area of the same sample F2.

\begin{figure}
\includegraphics[height=0.4\columnwidth]{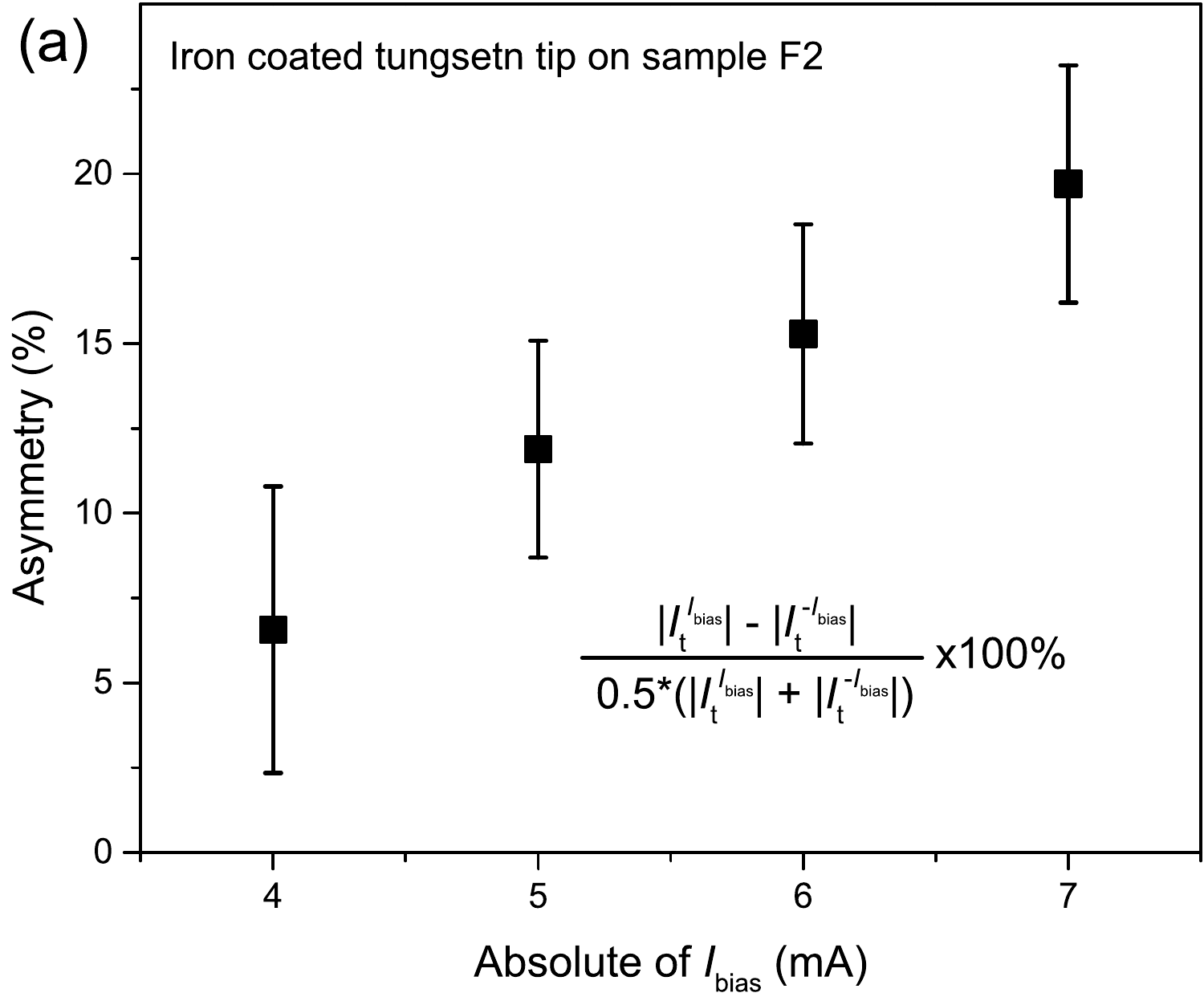}\hfill
\includegraphics[height=0.4\columnwidth]{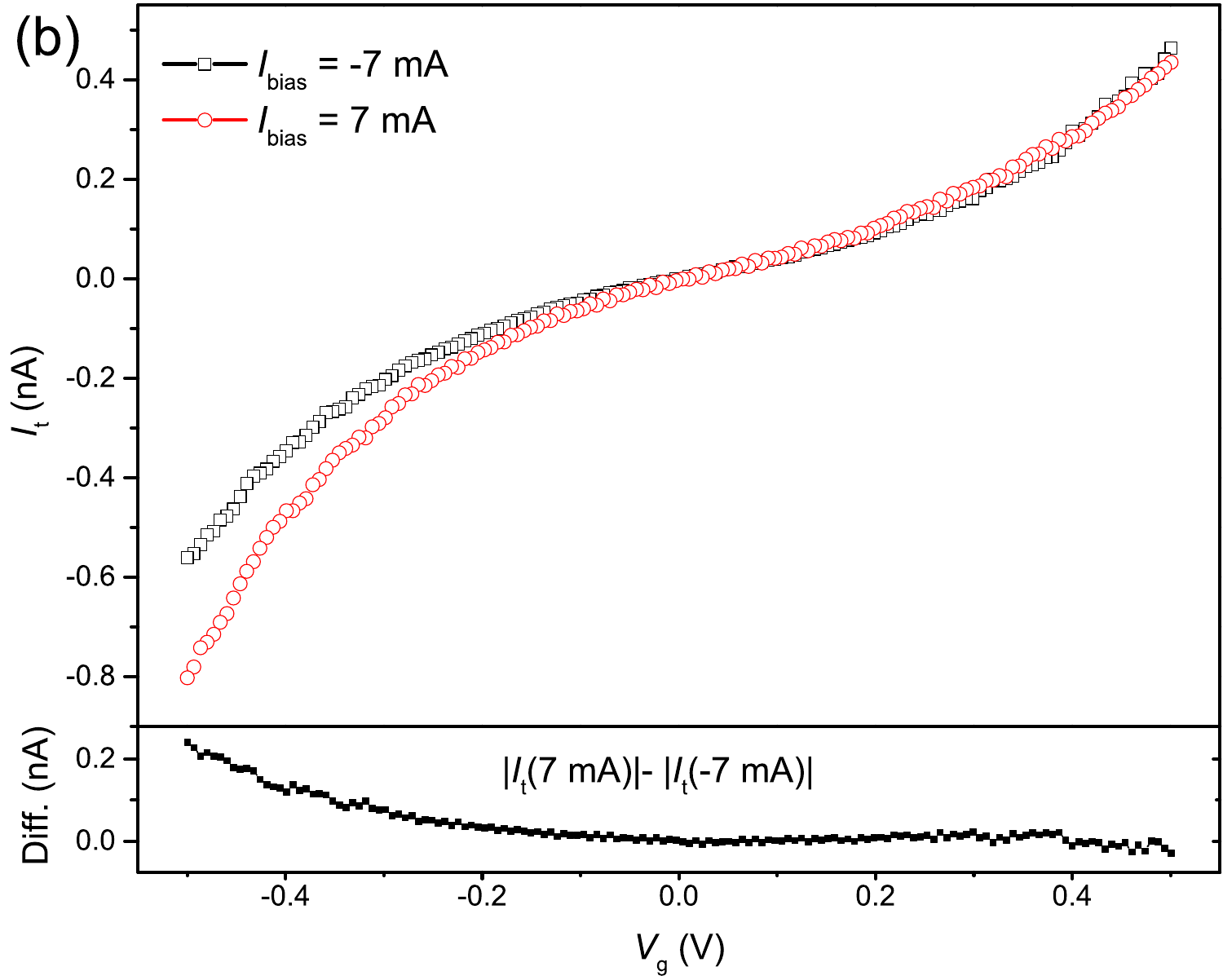}\vspace{8pt}
\includegraphics[width=0.49\columnwidth]{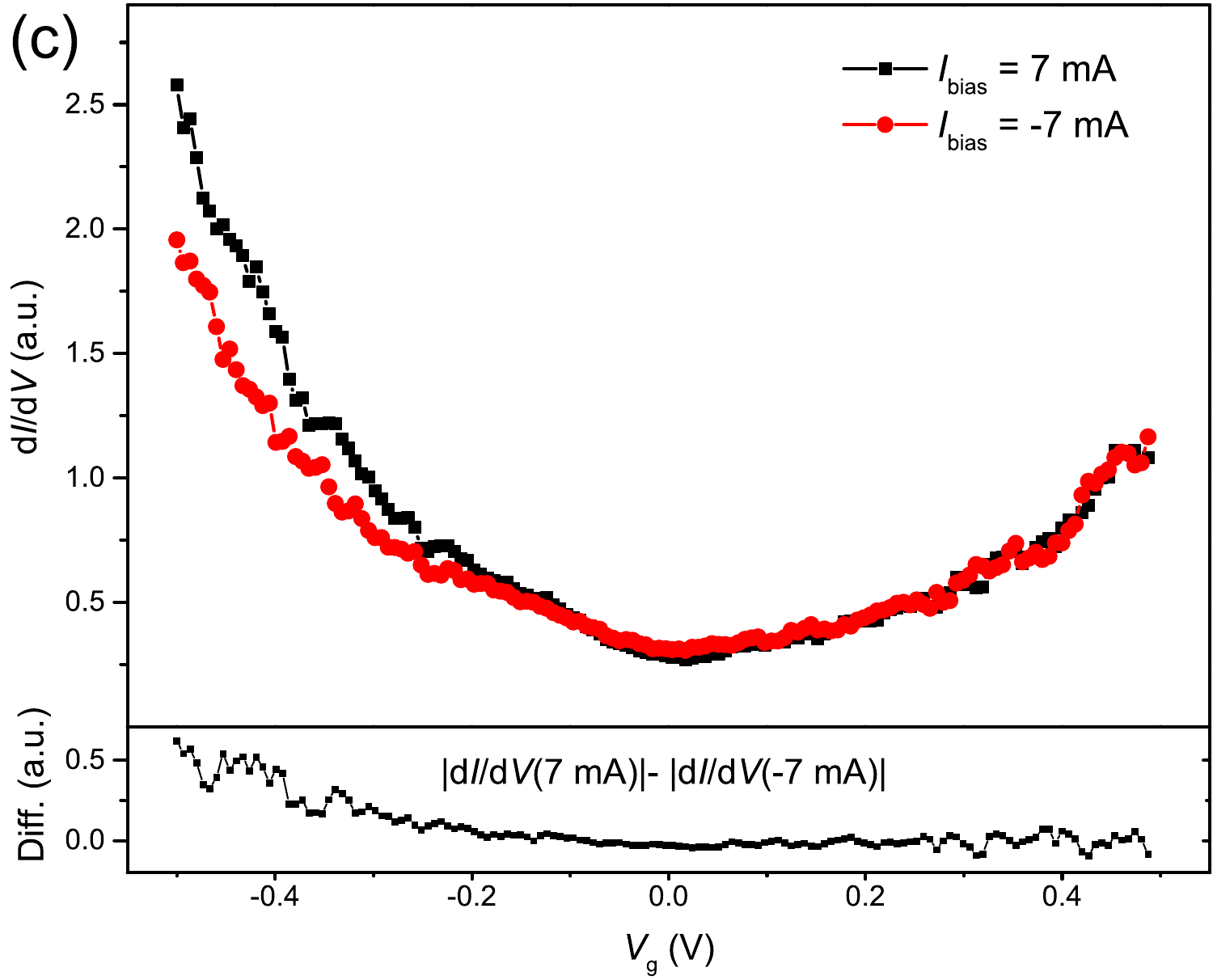}%
\caption{\label{fig5} (a) Tunneling current asymmetry with respect to different directions of the bias current measured with an iron-coated tip on tungsten film F2 as a function of the bias current magnitude. The asymmetry was calculated by using the equation shown in the figure, where $|I_{t}^{(Ibias)}|$ and $|I_{t}^{(-Ibias)}|$ are the magnitudes of tunneling currents measured at positive and negative directions of the bias current, respectively. The applied tunneling voltage is -0.5 V in these measurements. (b) and (c) $I_{t}-V_{g}$ and \didv\ curves measured with the iron-coated tip for different directions of a 7 mA bias current, respectively.}
\end{figure}

By further measurements with iron-coated tips, we confirmed the presence of spin-polarized electrons on the surface of the current-carrying tungsten film, as would be expected due to the SHE. A change in the bias current direction results in a change in the direction of electron spin-polarization induced by the SHE at the tungsten film surface. In the experiments with tungsten tips, no asymmetry in the tunneling current with respect to a change in the direction of the bias current flow through the tungsten film was observed (see Supplemental Material Fig. S6\cite{Supplemental}). This is consistent with the fact that tungsten tips do not possess a spin-dependent density of states (DOS). To sense the effect caused by a change in the direction of the bias current, iron-coated (spin-polarized) STM tips were used. These tips were prepared by depositing a thin layer of iron ($\sim18$ nm) onto tungsten tips \textit{in-situ}. Remarkably, the STM experiments with spin-polarized tips reveal that, when electrons tunnel from the film to the tip, there occurs a clear and pronounced asymmetry in tunneling currents caused by the reversal of the bias current direction through the film. This is illustrated in Fig.~\ref{fig5}a which also shows this asymmetry as a function of bias current magnitude. The observed asymmetry is based on the spin-dependent DOS of the iron-coated tungsten tip. To reveal the spin-dependent DOS of the iron-coated tungsten tip, STM based \didv -measurements were carried out by using a lock-in amplifier to experimentally differentiate the simultaneously measured $I-V$ signal. The use of bias current pulses was found to be detrimental to $I-V$ and \didv -measurements due to the displacement currents and the change in the tip-sample separation. To circumvent this problem, we performed the measurements after thermal equilibrium was reached. This was achieved by subjecting the tungsten film to a desired bias current flow for $\sim6-8$ hours. The results are presented in Fig.~\ref{fig5}b and~\ref{fig5}c. It is clear from these figures that for a negative tunneling voltage when electrons tunnel from film to tip, there are asymmetries in the $I-V$ and \didv\ signals upon changing the bias current direction. The asymmetry in $I-V$ curves shows that the magnitude of the tunneling current is larger with a positive bias current, which agrees with the pulsed bias current measurement shown in Fig.~\ref{fig5}a. The asymmetry in \didv\ signals reveals the spin-dependent DOS of the iron-coated tip. Again, these results demonstrate that the tunneling current asymmetry with respect to opposite directions of the bias current can be attributed to a change in the direction of the SHE-induced spin-polarization detected by using a spin selective tip. The Fig.~\ref{fig5}b also reveals the tunneling voltage polarity dependent asymmetry consistent with Fig.~\ref{fig4}a.
\section{Conclusion}
The reported experimental results reveal the ability of using STM for local sensing of spin-polarized electron accumulation due to the SHE in current-carrying tungsten samples. Two different types of SHE-induced tunneling current asymmetries were observed by using spin-polarized and non-spin polarized tips upon reversing the bias current direction and tunneling voltage polarity, respectively. These two asymmetries result from different physical origins but both reveal the presence of spin-polarized electrons at the tungsten surface. It is demonstrated that due to the ultra-high sensitivity of the STM to changes in tunneling conditions, the SHE can be sensed by using much smaller bias current densities than those used with other analysis methods \cite{Pai}. 


\providecommand{\noopsort}[1]{}\providecommand{\singleletter}[1]{#1}%
\pagebreak
\textbf{Supplementary Material 1: Scanning tunneling microscopy (STM) images of the fabricated tungsten films}

Representative STM images obtained on the fabricated tungsten films with tungsten tips and iron coated tungsten tips are presented in Fig. S1a-f. The surface morphologies of sample F1, F2 and F3 are similar. Sample F4 shows a much larger grain size than the other three tungsten samples, which is due to its much higher film thickness. Fig. S1f shows the topography of the pre-patterned 1 μm thick gold contacts. Histograms of the grain areas for the samples F1 and F2 are presented in Fig. S1g. The similarity of these histograms indicates good control over the film deposition process.\\
\textbf{Supplementary Material 2: $I_{t}-V_{s}$ curves measured from the potentiometry method to determine the bias-current-induced $V(x, y)$}

In order to compensate the bias-current-induced gap voltage offset $V(x, y)$, a scanning tunneling microscopy (STM) based potentiometry technique was developed and implemented as shown in Fig. S2a and the measured $I_{t}-V_{s}$  curves by using the potentiometry technique are shown in Fig. S2b. The potentiometry measurement started by parking the STM tip at the desired location (x, y). Then, the feedback loop of the STM system was turned off to freeze the sample and tip at their current positions, i.e. no more mechanical movements. In the meantime, the tunneling current It was being measured with respect to the change of the source voltage Vs. Note, in this experiment we used the terminal of the sample which is connected to the voltage source as the reference point for the surface potential.\\
\textbf{Supplementary Material 3: Displacement current observed at the edge of a bias current pulse} 

The shape and amplitude of the spikes in the tunneling current at the edges of the bias current pulse depend on the edge slope as shown in Fig. S3. This indicates that the observed spikes are due to the displacement currents caused by the edge of the bias current pulse and the resulting transient in the tunneling voltage due to the finite response time of the voltage compensation circuit.\\
\textbf{Supplementary Material 4: Tip-sample distance change due to the bias current}
 
Figure S4a shows the measured tunneling current in the presence of $\sim1$ ms bias current pulses with amplitudes from 5 to 10 mA. It is clear from this figure that the tunneling current increases in time. Furthermore, this increase is found to be exponentially dependent on the amplitude of the bias current pulse, as revealed by the inset logarithm plot of the tunneling current as a function of the bias current. The observed increase in the tunneling current can be attributed to a reduction in the tip-sample separation. An atomic force microscopy (AFM) study was performed on tungsten film \textit{in-situ} to confirm the change in the tip-sample distance in the presence of the bias current pulses. The feedback was turned off during the AFM measurement as in the STM experiments. The results are presented in Fig. S4b. In this figure, the measured resonance frequency of the AFM cantilever shows a clear downward shift (negative df) in the presence of the bias current pulse. Since the AFM tip was stabilized on the attractive part of the force-distance curve, this shift reflects an increase in the attractive force gradient acting on the tip, which is evidence for the tip-sample distance reduction in the presence of bias current.\\
\textbf{Supplementary Material 5: Asymmetry as a function of bias current measured with a tungsten tip}

Figure S5a shows the measured asymmetry as a function of bias current with reversed sequences of the tunneling voltage polarity measured on sample F1 with a tungsten tip. For the sequence 1, the applied tunneling voltage was first negative and then switched to positive as also shown in Fig. 2b. The sequence of the tunneling voltage polarity was reversed in sequence 2, i.e. positive polarity for the first bias current pulse and negative polarity for the second pulse. The observed asymmetry as a function of the bias current is practically the same for either sequence. This result implies that any possible change in the tunneling gap due to a residual thermal expansion of the sample can be precluded as a cause of the observed tunneling current asymmetry. Figure S5b shows asymmetries measured at different locations on the sample F1. The locations exhibit a similar saturation pattern at a bias current around 7 mA. The variation in the asymmetry may be attributed to the non-uniformity in the local expression of the SHE due to the local surface structure of the fabricated tungsten film. \\
\textbf{Supplementary Material 6: Tunneling current measured with respect to a change in the direction of the bias current using a tungsten tip}

The tunneling current shows no noticeable difference with respect to a change in the direction of the bias current when measured with a tungsten tip. 

\begin{figure}
\includegraphics[width=1\columnwidth]{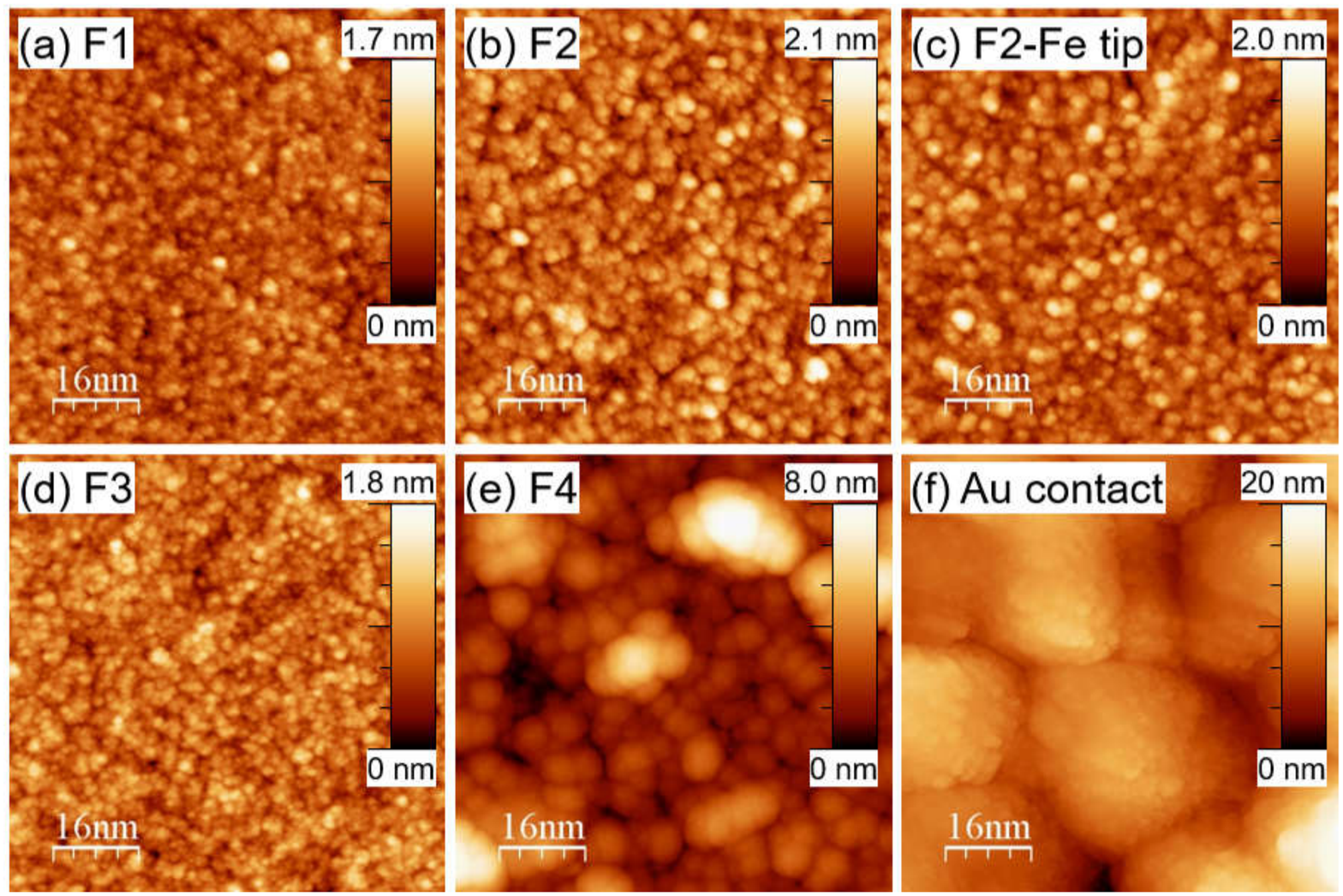}\vspace{8pt}
\includegraphics[height=0.4\columnwidth]{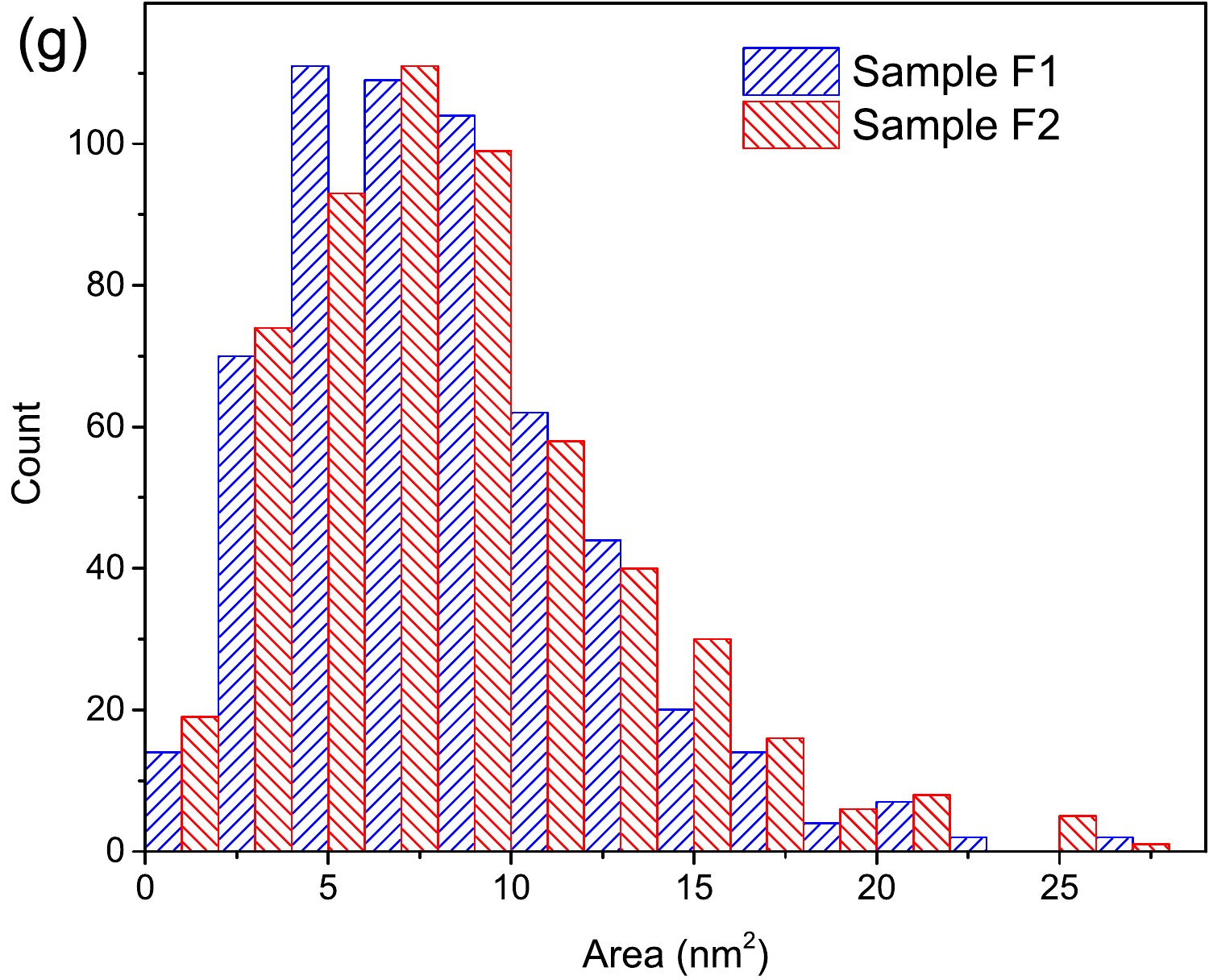}

Figure S1. STM images obtained with tungsten tips on (a) sample F1, (b) sample F2, (d) Sample F3, (e) Sample F4, and (f) gold contact of sample F2. (c) STM image of sample F2 obtained with an iron coated tungsten tip. (g) Grain area histograms of STM images (a) and (b).
\end{figure}

\begin{figure}
\includegraphics[width=0.49\columnwidth]{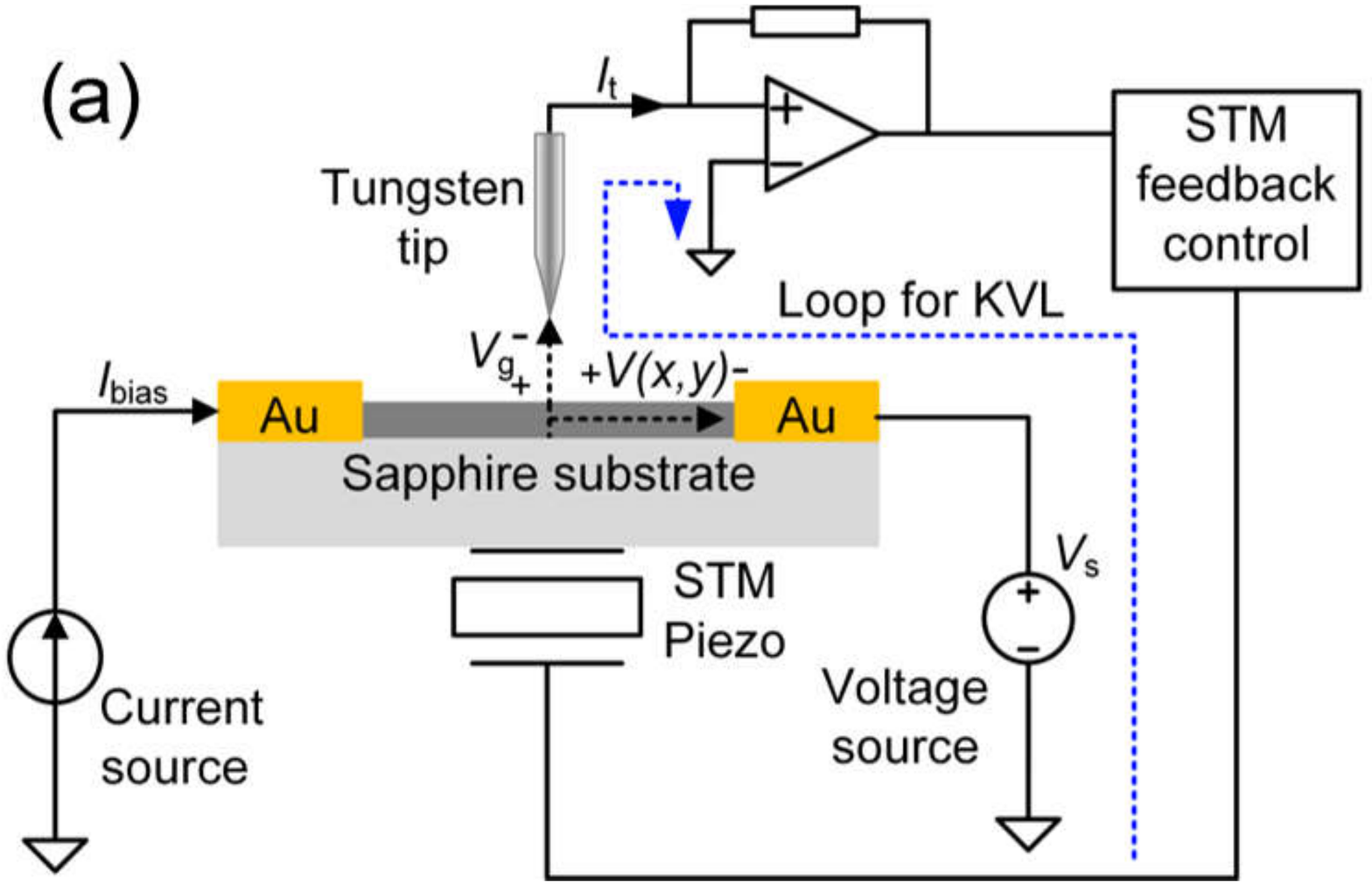}%
\includegraphics[width=0.49\columnwidth]{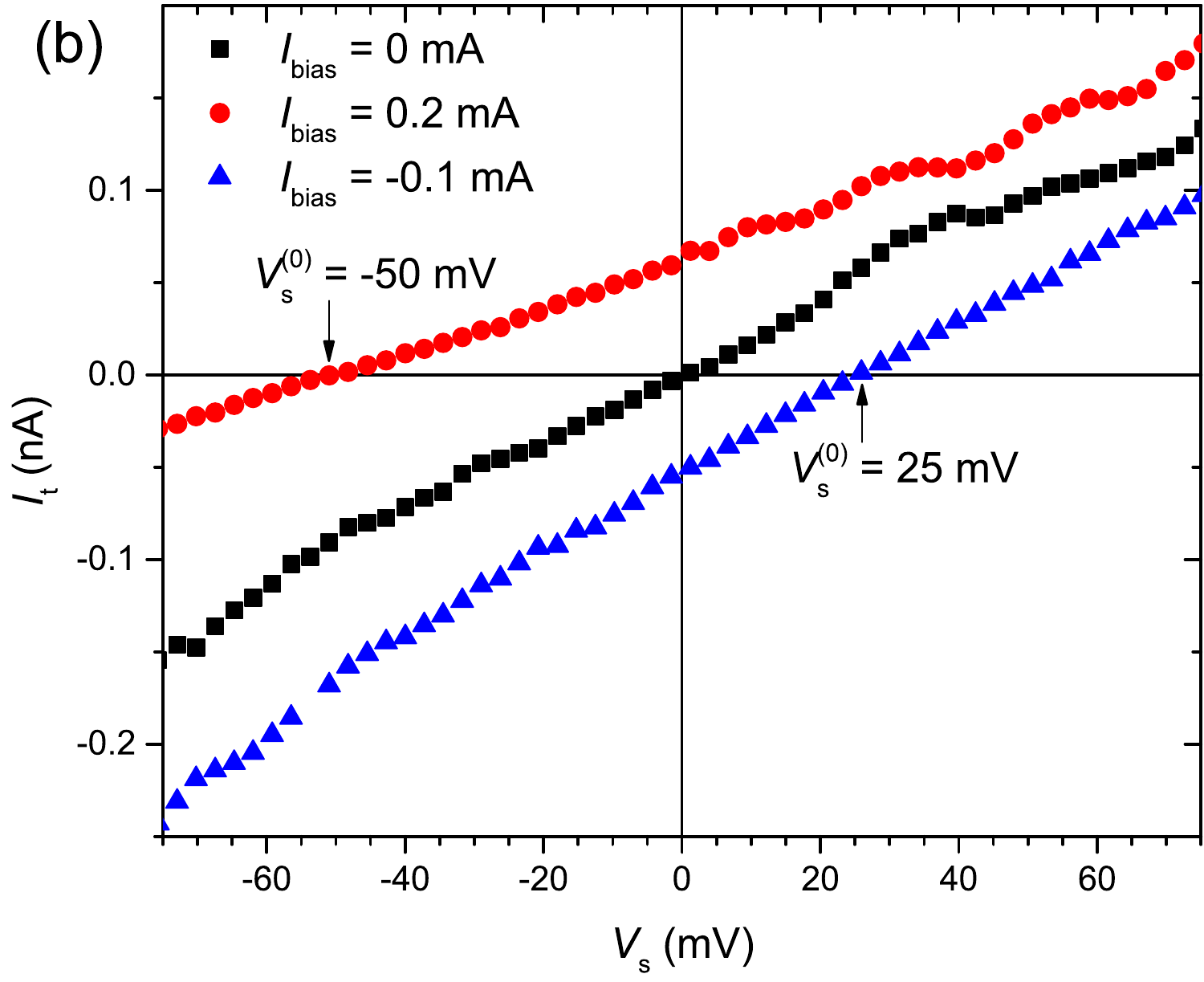}

Figure S2. (a) Schematic diagram for the experimental setup with the indicated Kirchhoff’s voltage law (KVL) loop for potentiometry measurements. (b) Typical $I_{t}-V_{s}$ curves and the corresponding values of $V(x, y)$ obtained at various $I_{bias}$.
\end{figure}
 
\begin{figure}
\includegraphics[width=0.6\columnwidth]{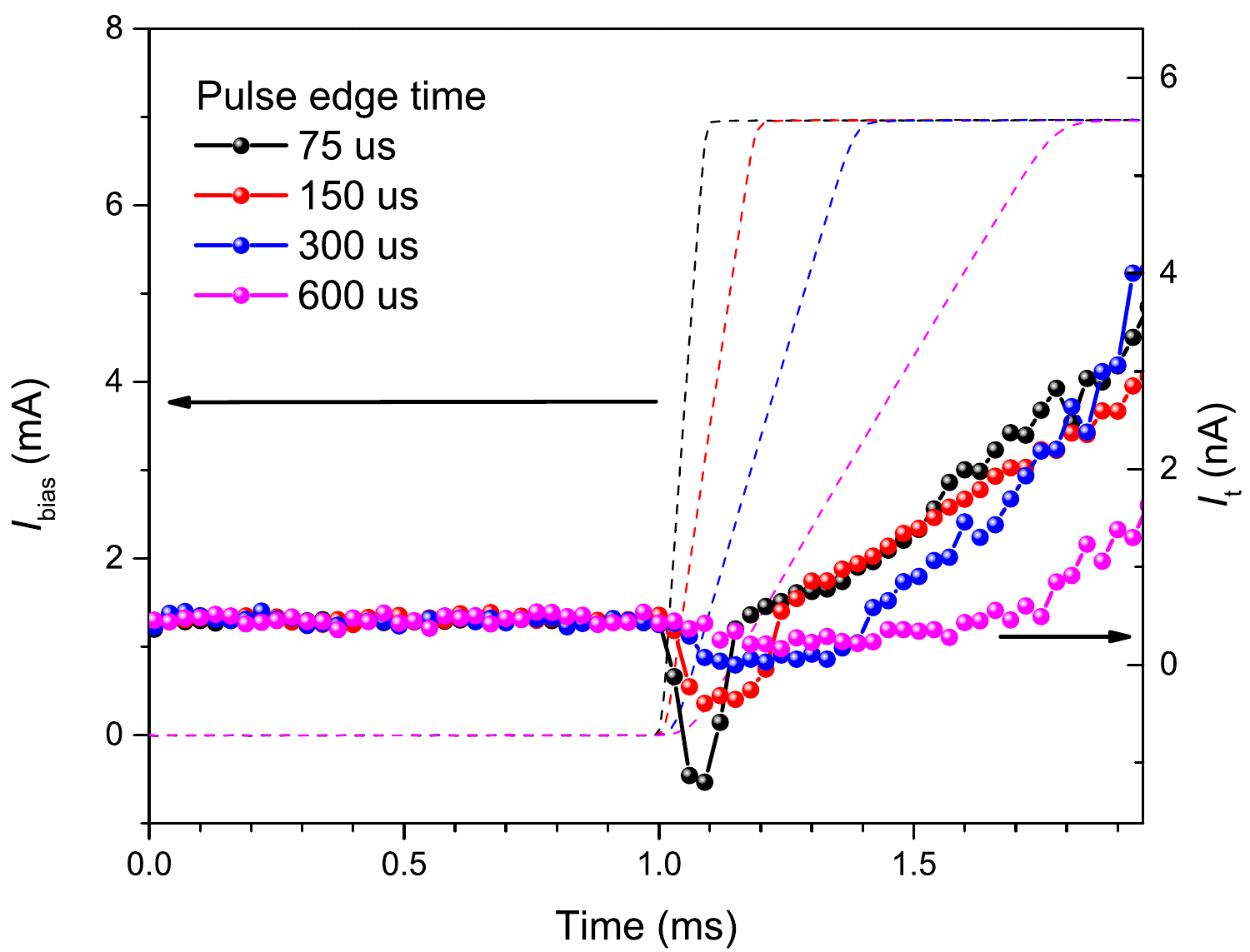}%

Figure S3. Tunneling current measured in the presence of 7 mA bias current pulses with varying pulse edge times. 
\end{figure}

\begin{figure}
\includegraphics[width=0.49\columnwidth]{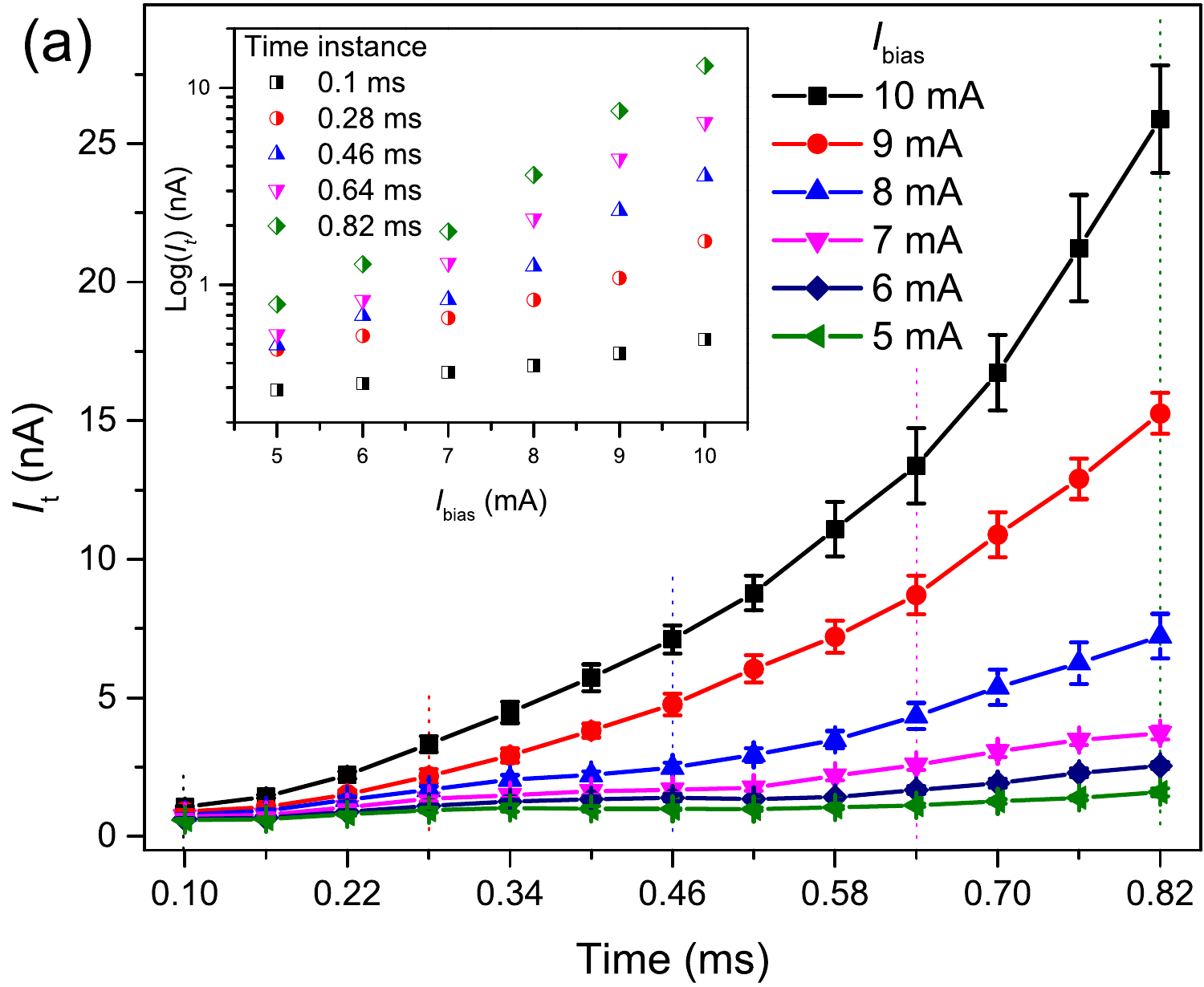}%
\includegraphics[width=0.49\columnwidth]{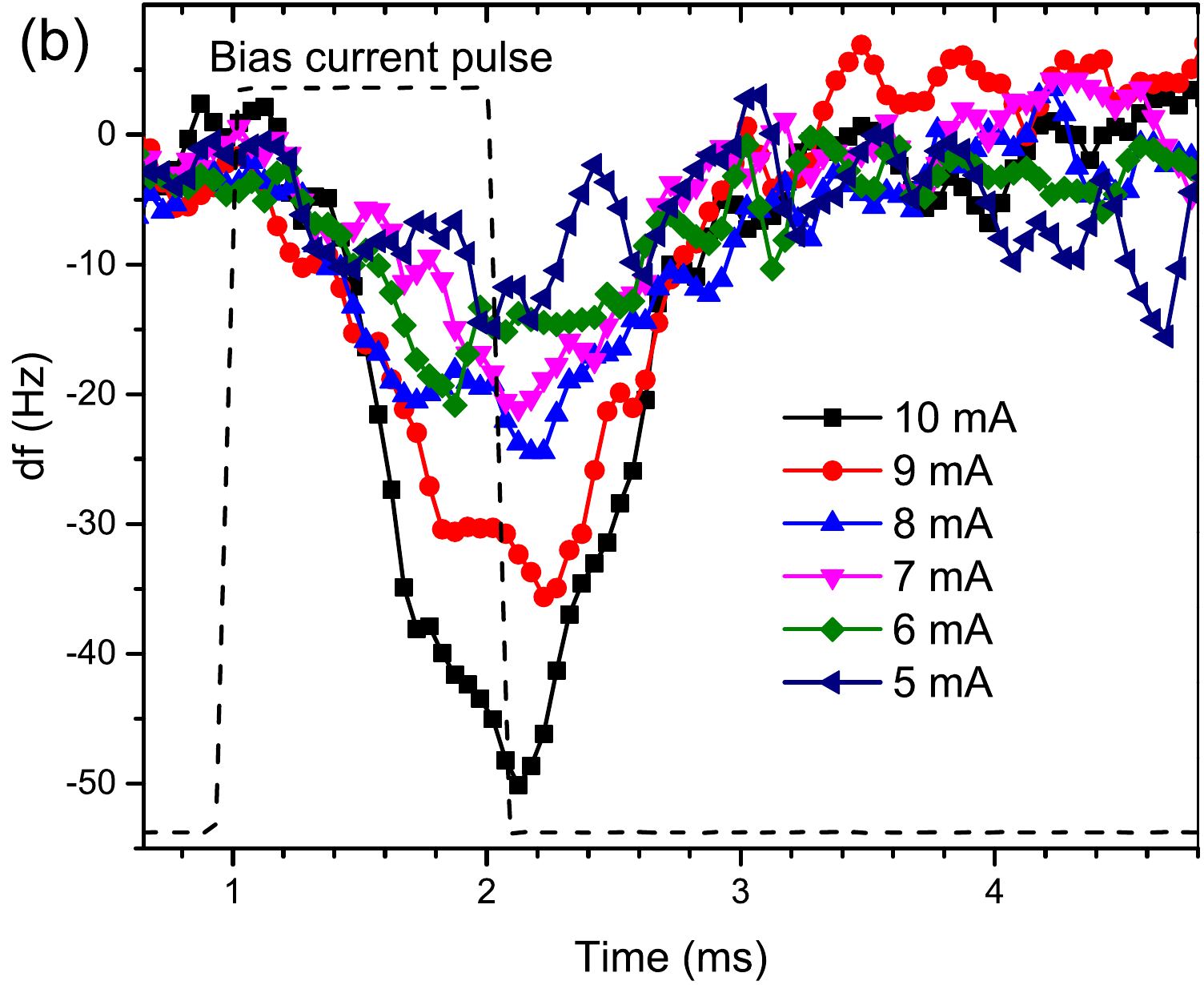}

Figure S4. (a) Tunneling currents in the presence of varying magnitudes of bias currents. The inset shows the tunneling current as a function of the bias current measured at various time instances. (b) Atomic force microscopy measurements on the tungsten film in the presence of varying bias current pulses.
\end{figure}

\begin{figure}
\includegraphics[width=0.49\columnwidth]{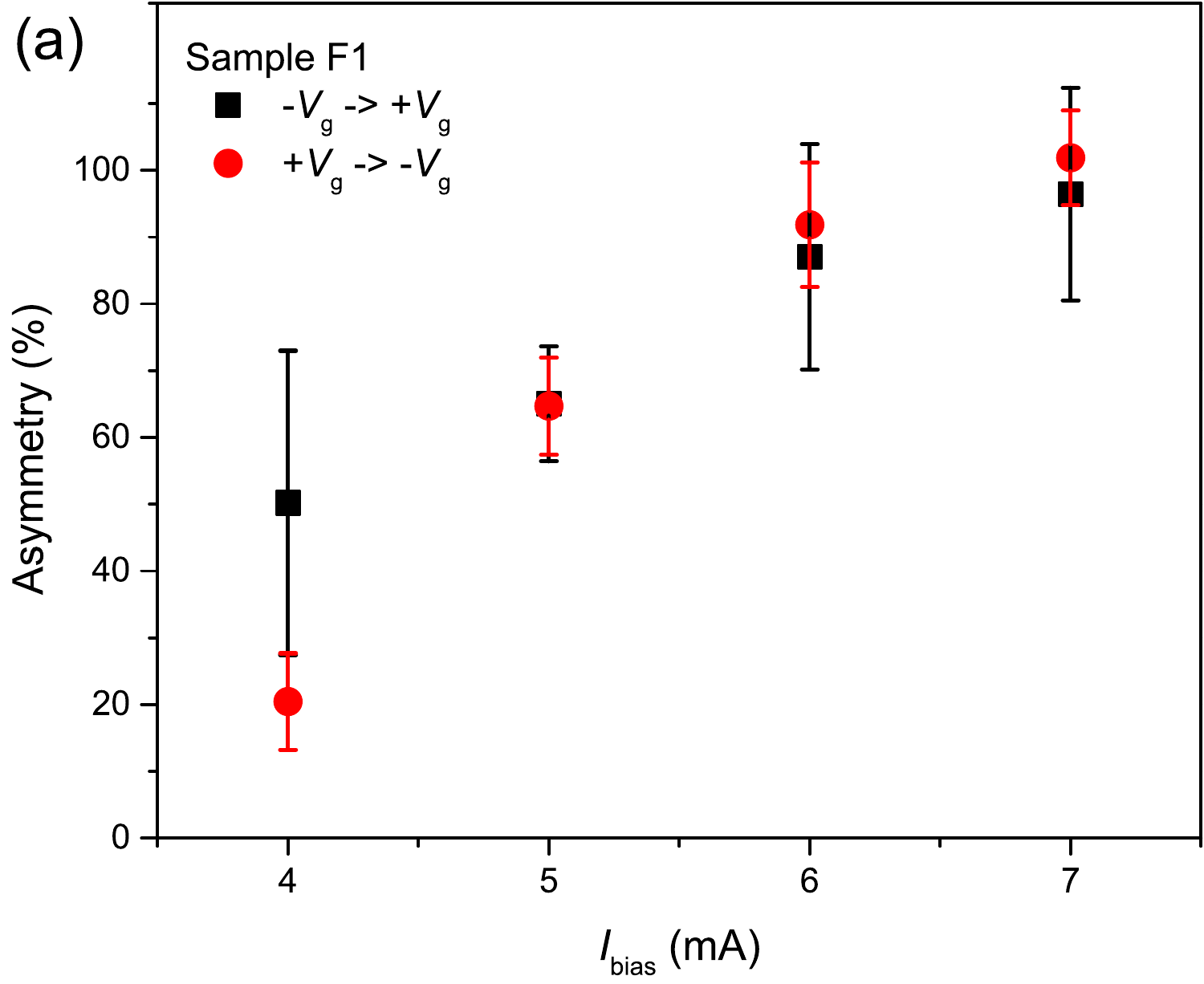}%
\includegraphics[width=0.49\columnwidth]{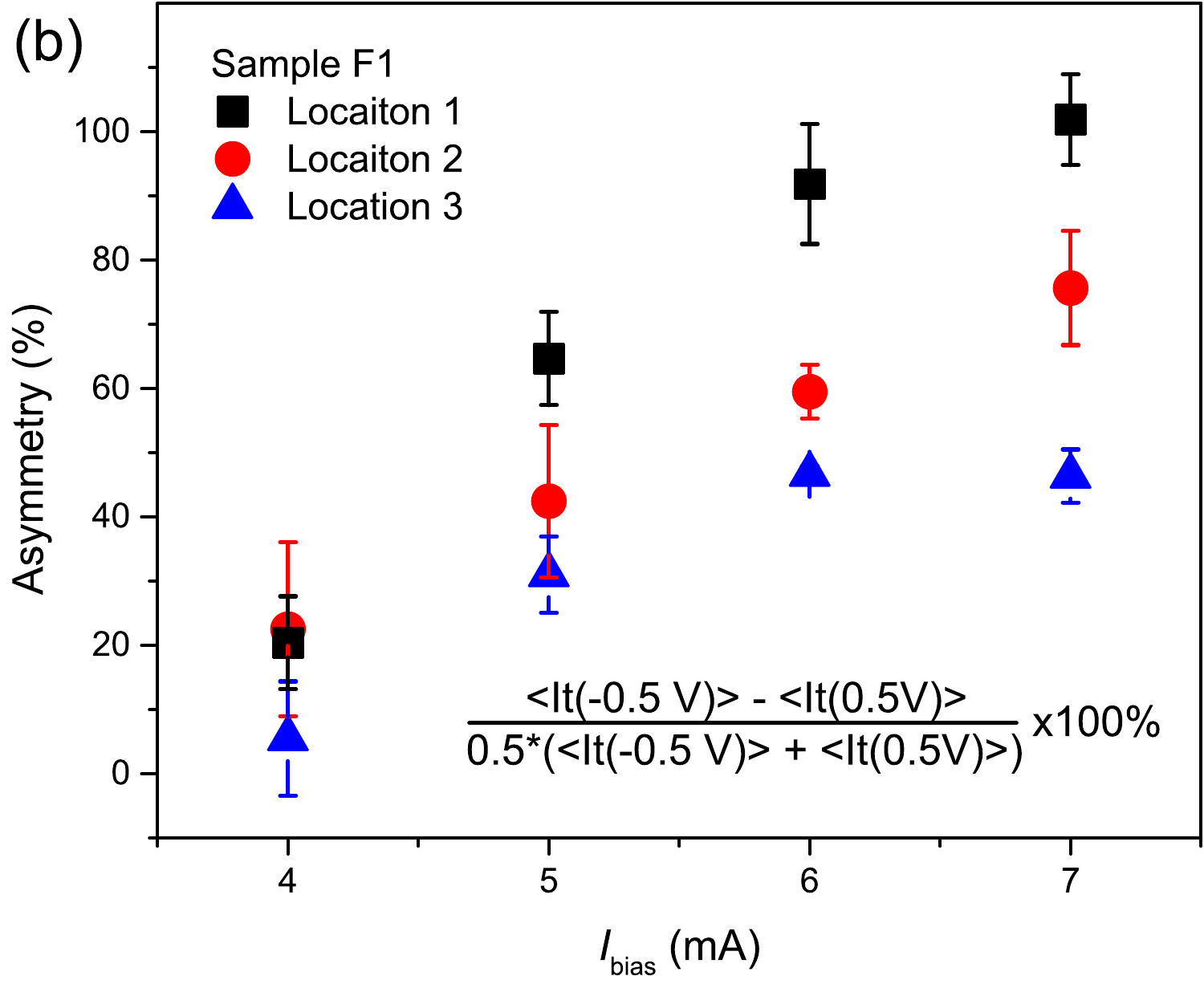}

Figure S5. (a) Asymmetry in tunneling currents as a function of bias current measured by opposite sequences of the polarity of the applied tunneling voltage. (b) Asymmetry as a function of bias current measured at three different locations on sample F1 by using a tungsten tip. 
\end{figure}

\begin{figure}
\includegraphics[width=0.7\columnwidth]{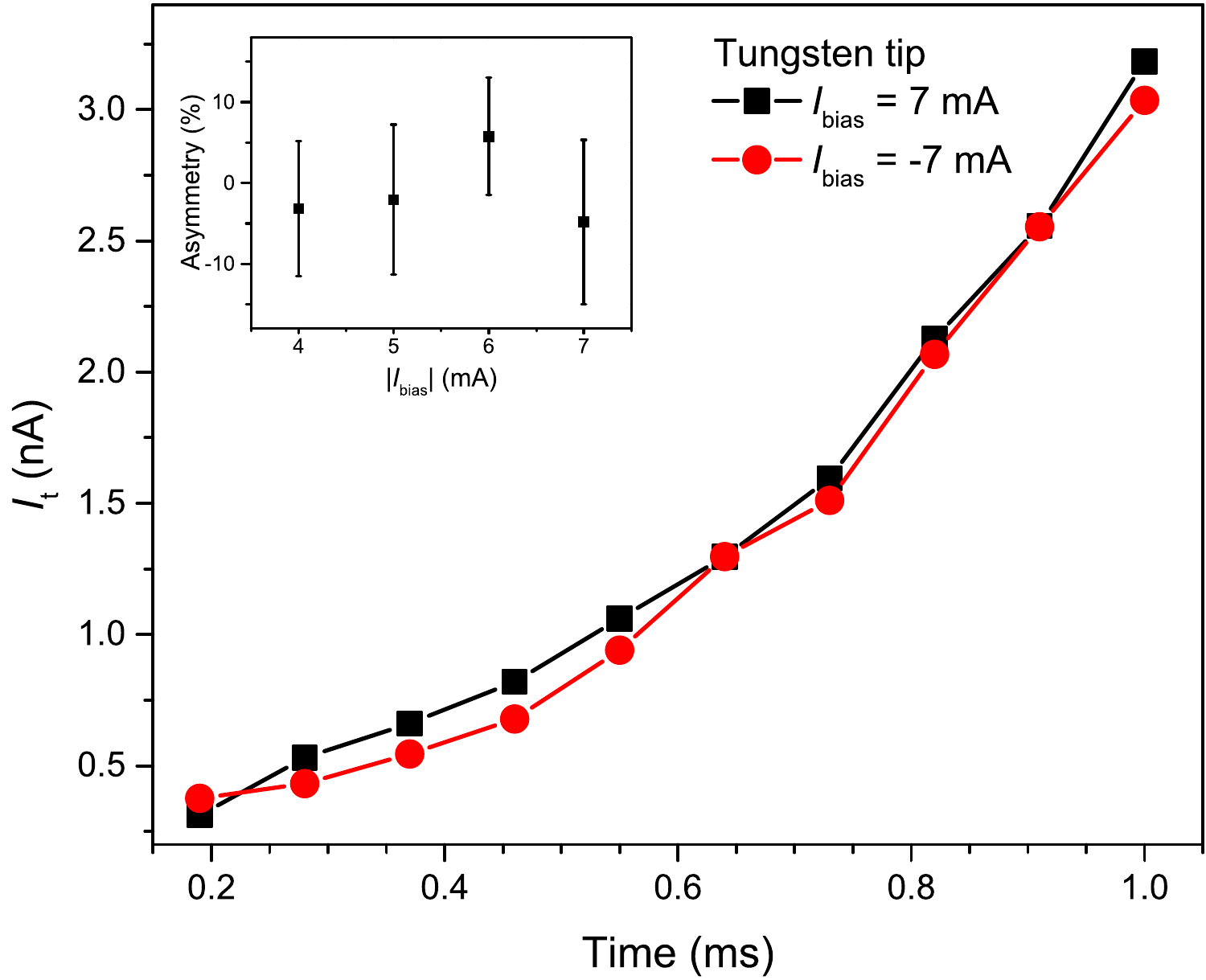}%

Figure S6. Tunneling currents as a function of time with respect to different directions of a 7 mA bias current measured by using a tungsten STM tip. Inset: the normalized tunneling current asymmetry with respect to different directions of the bias current measured in the tungsten tip-tungsten film system as a function of the bias current magnitude.  
\end{figure}
\end{document}